\documentclass[reprint,aps,prl,showpacs,twocolumn,letterpaper]{revtex4}
\usepackage{amssymb,amsfonts,amsmath}
\usepackage{times}
\usepackage{float}
\usepackage{graphicx}

\usepackage{color}

\begin{document}

\title{Decomposition Products of Phosphine Under Pressure: PH$_2$ Stable and Superconducting?}
\author{Andrew Shamp}
\author{Tyson Terpstra}
\author{Tiange Bi}
\author{Zackary Falls}
\author{Patrick Avery}
\author{Eva Zurek}\email{ezurek@buffalo.edu}
\affiliation{Department of Chemistry, State University of New York at Buffalo, Buffalo, NY 14260-3000, USA}

\begin{abstract}
Evolutionary algorithms (EA) coupled with Density Functional Theory (DFT) calculations have been used to predict the most stable hydrides of phosphorous (PH$_n$, $n=1-6$) at 100, 150 and 200~GPa. At these pressures phosphine is unstable with respect to decomposition into the elemental phases, as well as PH$_2$ and H$_2$. Three metallic PH$_2$ phases were found to be dynamically stable and superconducting between 100-200~GPa. One of these contains five formula units in the primitive cell and has $C2/m$ symmetry (5FU-$C2/m$). It is comprised of 1D periodic PH$_{3}$-PH-PH$_{2}$-PH-PH$_{3}$ oligomers. Two structurally related phases consisting of phosphorous atoms that are octahedrally coordinated by four phosphorous atoms in the equatorial positions and two hydrogen atoms in the axial positions ($I4/mmm$ and 2FU-$C2/m$) were the most stable phases between $\sim$160-200~GPa. Their superconducting critical temperatures ($T_c$) were computed as being 70 and 76~K, respectively, via the Allen-Dynes modified McMillan formula and using a value of 0.1 for the Coulomb pseudopotential, $\mu^*$. Our results suggest that the superconductivity recently observed by Drozdov, Eremets and Troyan when phosphine was subject to pressures of 207~GPa in a diamond anvil cell may result from these, and other, decomposition products of phosphine. 
\end{abstract}

\maketitle

\section{Introduction}

The pressure variable can be used to synthesize materials with unique stoichiometries, and properties that would not be accessible otherwise \cite{Grochala:2007a,Zurek:2014i,Hermann:2014a,Hoffmann:2012a,Ma:2012a,Ma:2015b} resulting in chemically different bonding motifs that are not observed at ambient conditions\cite{Ma:2013a,Hoffmann:2011a}. Recent experimental work by Drozdov, Eremets and Troyan has shown that hydrides of the $p$-element sulfur are superconducting with a critical temperature, $T_c$, of 203~K at 150~GPa \cite{hns}. This work was inspired by first-principles calculations suggesting that at $P>$ 130~GPa H$_2$S would transform to a phase that has a $T_c$ of 80~K above 160~GPa \cite{Ma:2014a}. The much higher critical temperature observed experimentally led to the suggestion that under pressure hydrogen sulfide decomposes into a hydride with a stoichiometry that is not stable at 1~atm, and this new phase is responsible for the remarkably high $T_c$ \cite{hns}. Indeed, first principles-calculations \cite{Cui:2014a} confirmed that the $T_c$ (computed using the Allen-Dynes modified McMillan equation) of a novel $Im\overline{3}m$-H$_3$S phase matched exceedingly well with what 
was observed by Drozdov and co-workers. A plethora of theoretical calculations have verified that the H$_3$S stoichiometry is indeed the most stable at these pressures, analyzed the factors contributing to $T_c$, examined anharmonicity as well as the isotope effect, and computed $T_c$ for the related compounds selenium hydride and tellurium hydride \cite{Flores:arxiv,Papaconstantopoulos:2015,Bernstein:2015,Duan:2015,Errea:2015a,Akashi:2015,Zhong:arxiv,Ma:arvix}.  These studies highlight that new hydrogen-rich materials can be attained under pressure, some of which may have a $T_c$ higher than those previously thought possible for Bardeen-Cooper-Schrieffer (BCS)-type superconductors \cite{Ashcroft:2004a}, and have invigorated the search for high-temperature superconductivity in high-pressure hydrides. For example, recently the stability of hydrides with novel stoichiometries containing a group 15 element such as  P, As, and Sb have been studied computationally, and the superconducting properties of select systems were examined. \cite{Pnictogen}.

Recently, new exciting experiments have revealed that pressure-induced high-temperature superconductivity may be found in other hydrides of a $p$-block element. Drozdov, Eremets and Troyan measured superconductivity in phosphine, PH$_3$, which was liquefied in a diamond anvil cell and subsequently compressed \cite{Troyan:arxiv}. Resistance measurements revealed a $T_c$ of 30~K and 103~K at 83~GPa and 207~GPa, respectively. Structural information on the superconducting phases was not provided. The possible existence of a high temperature superconductor containing phosphorus and hydrogen motivated us to examine the structural landscape of these elements combined under pressure using the evolutionary algorithm \textsc{XtalOpt}. Similar to what was found for H$_n$S under pressure, it is likely that the observed superconducting properties do not arise from the hydride with the ``classic'' 1~atm stoichiometry, but instead decomposition products of phosphine, such as PH$_2$.
Between the pressures of 100-200~GPa phosphine is thermodynamically less stable than PH$_2$ and H$_2$, and several superconducting PH$_2$ phases are metastable with respect to solid hydrogen and phosphorous. The $T_c$ computed for the PH$_2$ structures via the Allen-Dynes modified McMillan equation are significantly larger than those expected for pure phosphorus, approaching the experimental values measured in ``phosphine''. Our findings therefore suggest that PH$_2$ may be another hydrogen-rich BCS-type superconductor.

\section{Computational Details}
\textit{A priori} crystal structure prediction calculations were carried out using the open-source evolutionary algorithm (EA) \textsc{XtalOpt} Release 8 and 9 \cite{Zurek:2011a,Zurek:2011f} that has previously been used to predict the structures of a variety of binary hydrogen-rich phases under pressure \cite{Zurek:2011d,Zurek:2011h,Zurek:2012a,Zurek:2012g,Zurek:2013f}. EA runs were carried out on the PH$_3$ stoichiometry  at 100, 150, and 200~GPa employing simulation cells with 1-6 formula units (FU) at 100~GPa and 2-3 FU at 150 and 200~GPa. In addition, structure searches were performed on the PH$_n$, $n$ = (1,2,4-6), systems at 100, 150, and 200~GPa using cells with 2-3 formula units, unless otherwise noted in Tables S1-S6 in the Supplementary Information (SI). Duplicate structures were detected via the \textsc{XtalComp} \cite{Zurek:2011i} algorithm. The lowest enthalpy structures from each search were relaxed in a pressure range from 100-200~GPa.

Geometry optimizations and electronic structure calculations were performed by using density functional theory as implemented in the Vienna \textit{Ab-Initio} Simulation Package (VASP) versions 5.2 and 5.4.1 \cite{Kresse:1993a}, with the gradient-corrected exchange and correlation functional of Perdew-Burke-Ernzerhof (PBE) \cite{Perdew:1996a}. The projector augmented wave (PAW) method \cite{Blochl:1994a} was used to treat the core states, and a plane-wave basis set with an energy cutoff of 700~eV was employed. The H 1$s$ and P 3$s$/3$p$ electrons were treated explicitly in all of the calculations, using the POTCARs for PAW-PBE H and PAW-PBE P available in the potpaw-PBE.52.tar.gz file from the VASP repository. The $k$-point grids were generated using the $\Gamma$-centered Monkhorst-Pack scheme, and the number of divisions along each reciprocal lattice vector was chosen such that the product of this number with the real lattice constant was 30~\AA{} in the structure searches, and 40-50~\AA{} otherwise. The $k$-meshes and energy cutoffs used resulted in enthalpies that were converged to within 1~meV/atom. It is likely important to employ density functionals approximating the effects of van der Waals (vdW) interactions 
for molecular solids containing $p$-block elements near ambient pressure. However, it has been shown that at higher pressures the effect of vdW interactions for these types of systems becomes negligible. For example, by $\sim$40~GPa dispersion forces were expected to have a minimal effect on the structural parameters, and as a result the properties, of CO$_2$ \cite{Schwerdtfeger:2013a}. For this reason we expect the effect of vdW interactions on the results presented herein,  which were obtained between 100 and 200~GPa, to be small. The Nudged Elastic Band (NEB) method \cite{Sheppard:2012a} was used to construct a reaction pathway between the $I4/mmm$ and $C2/m$ PH$_2$ phases at 200 GPa.

Phonon calculations were performed using the Quantum Espresso (QE) \cite{QE-2009} program to obtain the dynamical matrix and electron-phonon coupling (EPC) parameters. In the QE calculations, the H and P pseudopotentials, obtained from the QE pseudopotential library, were generated by the method of Trouiller-Martins with 1$s^2$ and 3$s^2$3$p^3$ valence configurations, respectively, along with the PBE generalized gradient approximation. Plane-wave basis set cutoff energies were set to 80~Ry for all systems. The Brillouin-zone sampling scheme of Methfessel-Paxton using a smearing of 0.02 Ry and $22\times 22 \times 22$ \textit{k}-point grids were used for all 100 GPa calculations of the PH$_2$ 2FU-$C2/m$ and $I4/mmm$ structures. At 150 and 200 GPa we used \textit{k}-point grids of $24\times 24 \times 24$ and $30\times 30 \times 30$ for the PH$_2$ 2FU-$C2/m$ and $I4/mmm$ structures, respectively. For the PH$_2$ 5FU-$C2/m$ structure a $12\times 12 \times 12$ \textit{k}-point grid was used. Density functional perturbation theory as implemented in QE was employed for the phonon calculations. The EPC matrix elements were calculated using $2\times 2 \times 2$ \textit{q}-meshes for all of the structures at 100 GPa, as well as for PH$_2$ 5FU-$C2/m$ at 150~GPa, and the phosphorus structures at all pressures. At 150 and 200~GPa $4\times 4 \times 4$ and $5\times 5 \times 5$ \textit{q}-meshes were used for the PH$_2$ 2FU-$C2/m$ and $I4/mmm$ structures, respectively. The EPC parameter, $\lambda$, was calculated using a set of Gaussian broadenings in steps of 0.005 Ry from 0-0.300~Ry. The broadening for which $\lambda$ was converged to within 0.05 was between 0.015 and 0.040~Ry for all structures. The critical superconducting temperature, $T_c$, has been estimated using the Allen-Dynes modified McMillan equation \cite{Allen:1975a} as,
\begin{equation}
T_{c} = \frac{\omega_{\text{log}}}{1.2} \text{exp} \left[\frac{1.04(1+\lambda)}{\lambda - \mu^*(1+0.62\lambda)}\right]
\end{equation}
where $\omega_\text{log}$ is the logarithmic average frequency and $\mu^*$ is the renormalized Coulomb potential, often assumed to be between 0.1 and 0.2.

\section{Results and Discussion}
\
\subsection{PH$_n$: Pressure Induced Decomposition of PH$_3$}

Because the phases that PH$_3$ adopts under pressure are unknown, we have carried out evolutionary searches to predict the global minima configurations up to 200~GPa. A detailed analysis will be published elsewhere, but the coordinates for the most stable structures found at $P\ge100$~GPa are provided in the SI. Briefly, the 100~GPa phase is made up of layers of PH and hydrogen, whereas the 150 and 200~GPa structures consist of 1-dimensional (1D) PH$_{3}$-PH$_{3}$ networks that resemble those present in the most stable PH$_2$ structure at 100~GPa discussed below. The computed enthalpies of formation, $\Delta H_f$, of PH$_3$ with respect to the most stable structures of solid hydrogen \cite{Pickard:2007a,Pickard:2012a} and phosphorus \cite{Kikegawa:1983,Clark:2010,Akahama:1999,Akahama:2000,Jamieson:1963} showed that the classic 1~atm stoichiometry is not thermodynamically stable at the pressures considered herein, and potentially even at lower pressures. The $\Delta H_f$ for the reaction $\frac{3}{2}\text{H}_2(\text{s})+\text{P}(\text{s})\rightarrow \text{PH}_3$ were computed to be 58.0, 45.2, and 42.6 meV/atom at 100, 150, and 200 GPa, respectively. 

\begin{figure}
\begin{center}
\includegraphics[width=0.7\columnwidth]{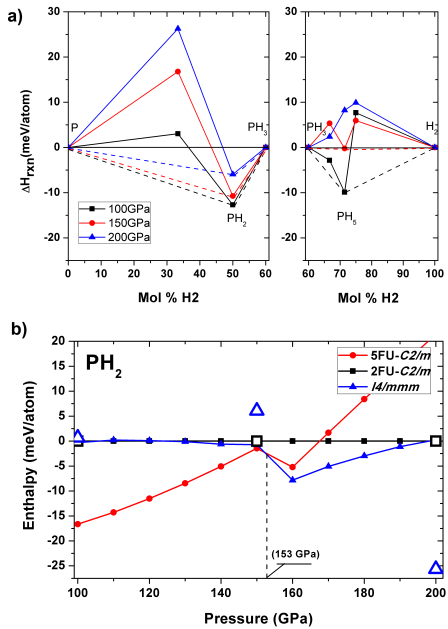}
\end{center}
\caption{(a) $\Delta H_\text{rxn}$ for the reaction: (left) $\text{PH}_n\rightarrow (\frac{3-n}{3})\text{P}+(\frac{n}{3})\text{PH}_3$ (where $n=0-3$) and (right) $\text{PH}_n\rightarrow (\frac{n-3}{2})\text{H}_2+\text{PH}_3$ (where $n=3-6$) at 100, 150, and 200~GPa. The convex hulls are given by the dashed lines. (b) The enthalpies as a function of pressure for the 5FU-${C2/m}$ and ${I4/mmm}$ phases of PH$_{2}$ relative to 2FU-${C2/m}$ between 100-200 GPa. The dashed line indicates the transition pressure below which the 5FU-${C2/m}$ phase has the lowest non-ZPE corrected enthalpy. The open symbols provide the relative enthalpies including the ZPE correction. At 150~GPa the ZPE-corrected enthalpy of 5FU-${C2/m}$ (not shown on the plot) is 45.8~meV/atom higher than that of the 2FU-${C2/m}$ structure.
\label{fig:Figure1}}
\end{figure}

Nonetheless, it is convenient to plot the $\Delta H_\text{rxn}$ of the hydrides of phosphorous, PH$_n$ with $n=1-6$, identified via our evolutionary searches for the reaction $\text{PH}_n\rightarrow (\frac{3-n}{3})\text{P}+(\frac{n}{3})\text{PH}_3$ (where $n=0-3$) and $\text{PH}_n\rightarrow (\frac{n-3}{2})\text{H}_2+\text{PH}_3$ (where $n=3-6$), as shown in the left and right hand sides of Fig.\ \ref{fig:Figure1}(a), respectively. For each of these plots a convex hull can be drawn, and the structures whose $\Delta H_\text{rxn}$ lie on the hull correspond to the most stable decomposition products of phosphine, or the most stable products arising from the reaction of phosphine with hydrogen. At 100, 150 and 200~GPa only PH$_2$ is thermodynamically preferred over P/PH$_3$, and the reaction PH$_3\rightarrow$PH$_2+\frac{1}{2}$H$_2$ is exothermic. At 100 and 150~GPa the reaction PH$_3+$H$_2\rightarrow$PH$_5$ is exothermic. 
Because PH$_5$ becomes unstable by 200~GPa we focused our analysis on PH$_2$ (the coordinates for PH$_5$ are provided in the SI). 

Evolutionary searches at 100, 150 and 200~GPa identified three unique low-enthalpy PH$_{2}$ phases, whose relative enthalpies are provided in Fig.\ \ref{fig:Figure1}(b). A $C2/m$ symmetry system whose primitive cell contained five formula units, 5FU-${C2/m}$, had the lowest enthalpy excluding zero point energy (ZPE) contributions below 153~GPa. A two formula unit cell with ${C2/m}$ symmetry (2FU-${C2/m}$) and an ${I4/mmm}$ structure became preferred at higher pressures. Even though these species are thermodynamically unstable relative to solid phosphorous and H$_2$ (by at least 38.7, 29.4, and 31.9 meV/atom at 100, 150 and 200~GPa), they are all dynamically stable at 100, 150 and 200 GPa. Moreover, the formation of PH$_n$ with $n = 4-6$ from PH$_2$ and H$_2$ at 100, 150 and 200~GPa, and the decomposition of PH$_2$ into H$_2$ and PH at 100 and 150~GPa is computed as being endothermic.

\begin{figure}[!ht]
\begin{center}
\includegraphics[width=0.7\columnwidth]{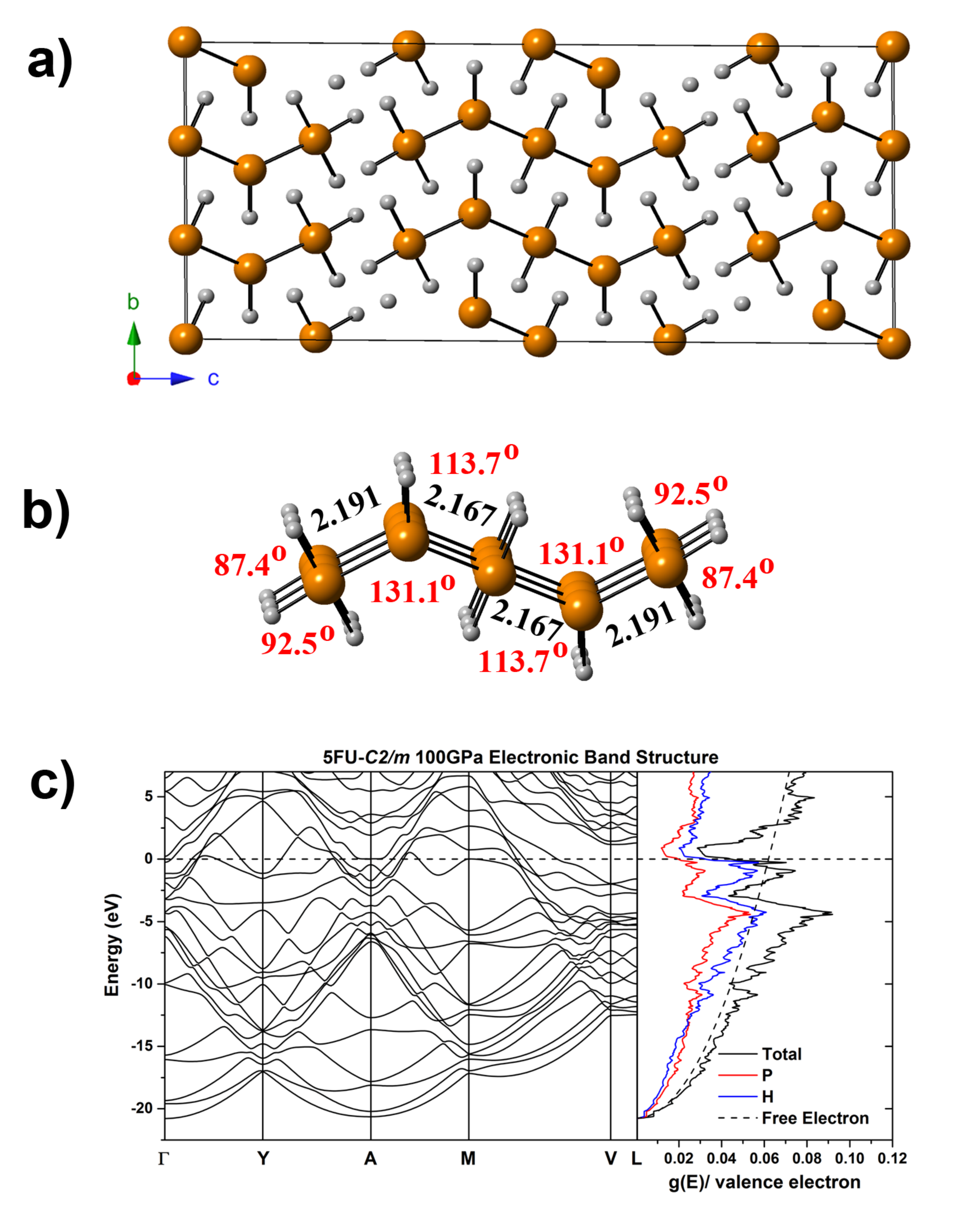}
\end{center}
\caption{(a) A 3$\times$3$\times$2 supercell of the 100 GPa 5FU-$C2/m$ structure. Phosphorous atoms are orange, hydrogen atoms are white. (b) A fragment of the $^1_\infty$[PH$_{3}$-PH-PH$_{2}$-PH-PH$_{3}$] motifs that comprise 5FU-$C2/m$. The values provided in black are P-P distances (\AA{}) and the red values are bond angles at 100~GPa. (c) Electronic band structure along with the total and site projected electronic DOS of 5FU-${C2/m}$ at 100~GPa. Because the PDOS for the H and P atoms were relatively independent upon the local atomic environment, we have not decomposed the PDOS into contributions from distinct H and P atoms. $E_F$ is set to zero.
\label{fig:Figure2}}
\end{figure}

The coordination and arrangement of the phosphorus atoms in 5FU-${C2/m}$ is quite different than in the 2FU-${C2/m}$ and ${I4/mmm}$ phases. The former consists of PH$_{3}$-PH-PH$_{2}$-PH-PH$_{3}$ oligomers extending along the $c$-axis, which are 1D periodic along the $a$-axis, $^1_\infty$[PH$_{3}$-PH-PH$_{2}$-PH-PH$_{3}$], see Fig.\ \ref{fig:Figure2}(a,b). The PH$_{3}$-PH-PH$_{2}$-PH-PH$_{3}$ unit possesses an inversion center of symmetry, with the H$_3$P-PH and HP-PH$_2$ distances measuring 2.191 and 2.167~\AA{}, respectively, at 100 GPa. The phosphorous atoms comprising the PH and PH$_2$ units are coordinated to four other phosphorous atoms forming a square net. Such structural motifs are not uncommon for compressed phosphorous, particularly in the pressure range between 100-200~GPa. 
For example, a number of the high pressure phases of elemental phosphorus, such as the A17 and A7 phases, have been described as Peierls distortions of a simple cubic lattice \cite{Hoffmann:1999, Clark:2010, Boulfelfel:2012}. Beyond the A7 phase of black phosphorus, a simple cubic lattice is adopted at $\sim$11.1~GPa that is stable up to 137~GPa \cite{Jamieson:1963,Kikegawa:1983,Akahama:1999}. After a relatively brief transition through a suspected incommensurate phase \cite{Ehlers:2004,Haussermann:2003}, simple cubic phosphorus transitions into a simple hexagonal phase \cite{Akahama:1999} followed by a body centered cubic phase adopted at pressures above 262~GPa \cite{Akahama:2000}.

The phosphorous atoms in the -PH$_3$ units at the end of the oligomer are nearly perfectly octahedrally coordinated to three hydrogen and three phosphorous atoms, with the H-P-H angles measuring 87.4 and 92.5$^{\circ}$ and the P-P-P angles measuring 90$^{\circ}$ at 100~GPa. Similarly, the phosphorous atoms comprising the -PH$_2$- units at the center of the motif also assume octahedral coordination with four phosphorous atoms on the equatorial and two hydrogen atoms on the axial positions. The phosphorous atoms within the -PH- segment of the oligomer, on the other hand, are nearly trigonal bipyramidal. The axial positions are occupied by two phosphorous atoms, and the equatorial positions are filled by two phosphorous and one hydrogen atom ($\measuredangle$PPP = 131.1$^{\circ}$, $\measuredangle$PPH = 113.7 and 115.2$^{\circ}$). At this pressure all of the P-H distances fall between 1.430-1.444~\AA{}, which is close to the typical experimental P-H bond length of 1.437~\AA{} \cite{Laby:1995} at 
1~atm. The P-P distances fall in the range of 2.167-2.191~\AA{} within the oligomer, and measure 2.151~\AA{} between oligomers, which is comparable to the distance we calculate for elemental phosphorous at this pressure, 2.149~\AA{}.

Fig.\ \ref{fig:Figure2}(c) shows that at 100~GPa 5FU-${C2/m}$-PH$_2$ is metallic. The occupied density of states (DOS) is nearly free electron like. But because the Fermi level, $E_F$, falls near a pseudogap the DOS at $E_F$, $g(E_F)$ = 0.045~eV$^{-1}$/valence electron, is lower than the value computed for a free electron gas with the same bandwidth, 0.063~eV$^{-1}$/valence electron. The metallicity is due primarily to hydrogen states with $s$-character and also phosphorous states with $p$-character. The occupied hydrogen and phosphorous states hybridize, indicative of covalent bonding. The electron localization function (ELF), provided in the SI, shows regions of high ELF along the P-P and P-H bonds, and no lone pairs are observed, as would be expected based upon valence-shell electron-pair repulsion (VSEPR) theory, which assumes that the geometry a molecule assumes minimizes the repulsion between valence shell electron pairs. A Bader analysis revealed electron transfer from phosphorous to hydrogen, with the average charge on the hydrogen atoms being -0.34 at 100~GPa. The charges on the phosphorous atoms in the -PH$_3$, -PH$_2$-, and -PH- units were computed as being +1.12, +0.70 and +0.25, respectively. At 150~GPa the average Bader charges become -0.37 (H), +1.15 (P in -PH$_3$), +0.75 (P in -PH$_2$-) and +0.30 (P in -PH-). Phonon calculations showed that this phase is dynamically stable at 150~GPa.

\begin{table*}
\begin{minipage}{\textwidth}
\centering
\caption{Phosphorus-phosphorus, phosphorus-hydrogen and hydrogen-hydrogen distances for the 2FU-$C2/m$ and $I4/mmm$ PH$_2$ phases. P-H(1)/H-H(1) are the first nearest neighbor, and P-H(2) are the second nearest neighbor distances. The average Bader charges per atom type are provided, along with the volume per unit cell.}
\begin{tabular}{c c c c c c c c c}
\hline
System & Pressure  & P-H(1)  & P-H(2)  & P-P     & H-H(1) & H       & P       & Volume  \\
       &   (GPa)   & (\AA{}) & (\AA{}) & (\AA{}) & (\AA{})& (${e}$) & (${e}$) & (\AA{}$^{
3}$/FU) \\  
\hline
2FU-$C2/m$ & 100 & 1.449 & 2.165 & 2.159 & 1.527 &  -0.26 & 0.53 & 13.885 \\
           & 150 & 1.436 & 2.013 & 2.111 & 1.497 &  -0.29 & 0.57 & 12.398 \\
           & 160 & 1.445 & 1.760 & 2.127 & 1.510 &  -0.27 & 0.55 & 12.025 \\
           & 200 & 1.439 & 1.693 & 2.093 & 1.500 &  -0.26 & 0.52 & 11.288 \\
$I_4/mmm$ & 100 & 1.449 & 2.163 & 2.157 & 1.528 &  -0.30 & 0.60 & 13.882 \\
          & 150 & 1.437 & 2.012 & 2.109 & 1.494 &  -0.28 & 0.56 & 12.401 \\
          & 160 & 1.435 & 1.989 & 2.101 & 1.490 &  -0.28 & 0.56 & 12.170 \\
          & 200 & 1.434 & 1.896 & 2.077 & 1.487 &  -0.26 & 0.52 & 11.359 \\
\hline
\end{tabular}
\label{tab:Distances}
\end{minipage}
\end{table*}

\begin{figure}
\begin{center}
\includegraphics[width=0.7\columnwidth]{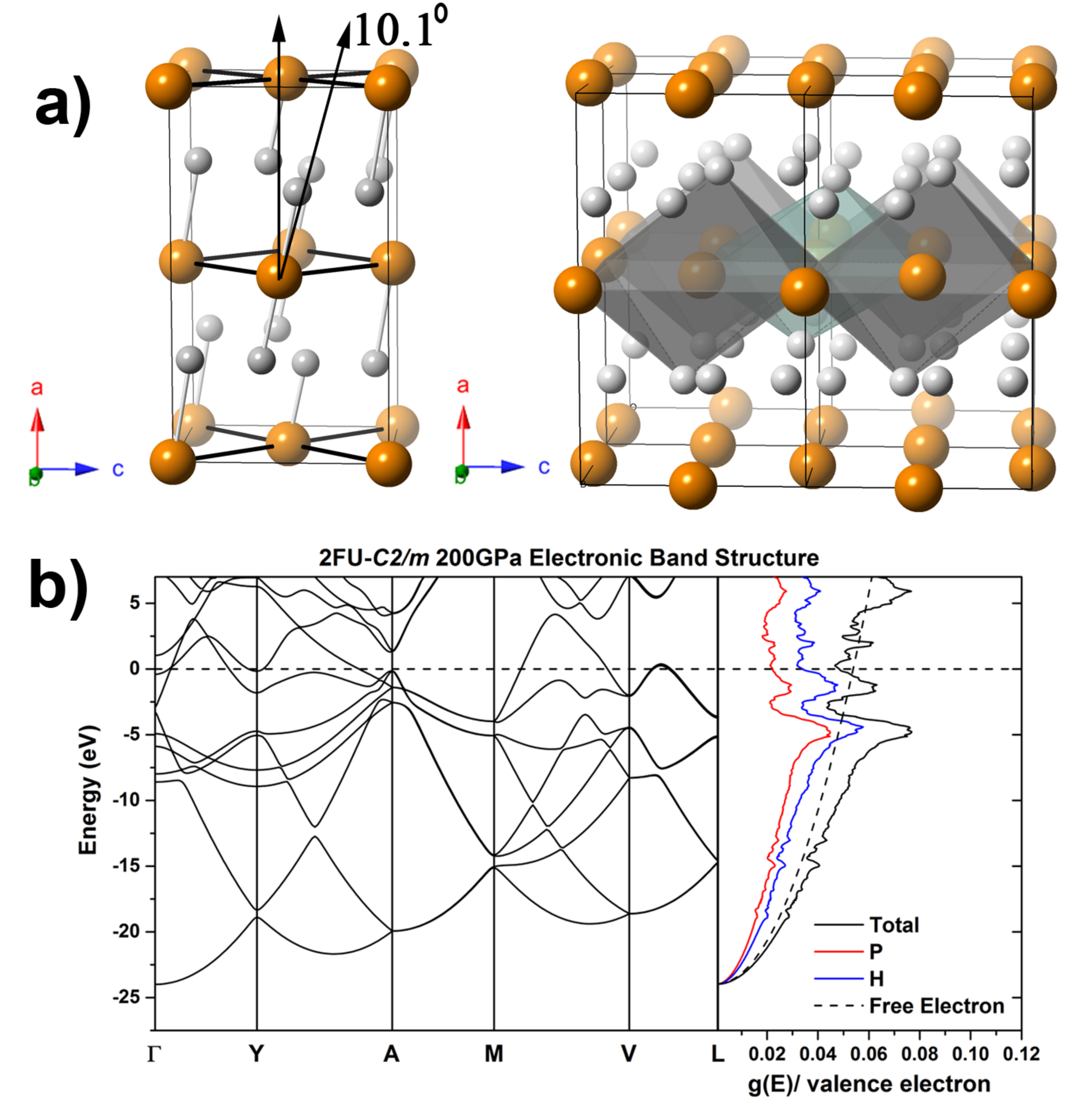}
\end{center}
\caption{(a) (left) A conventional unit cell and (right) a 1$\times$2$\times$2 supercell of the 200~GPa 2FU-${C2/m}$ phase. Phosphorous atoms are orange, hydrogen atoms are white. (left) The canting angle with respect to a line perpendicular to the plane containing the phosphorous atoms (10.1$^{\circ}$) is shown by the arrows. (right) The octahedra around the central plane of phosphorous atoms are highlighted by teal and gray. (b) Electronic band structure along with the total and site projected electronic DOS of 2FU-${C2/m}$ at 200~GPa. $E_F$ is set to zero.
\label{fig:Figure3}}
\end{figure}

\begin{figure}
\begin{center}
\includegraphics[width=0.7\columnwidth]{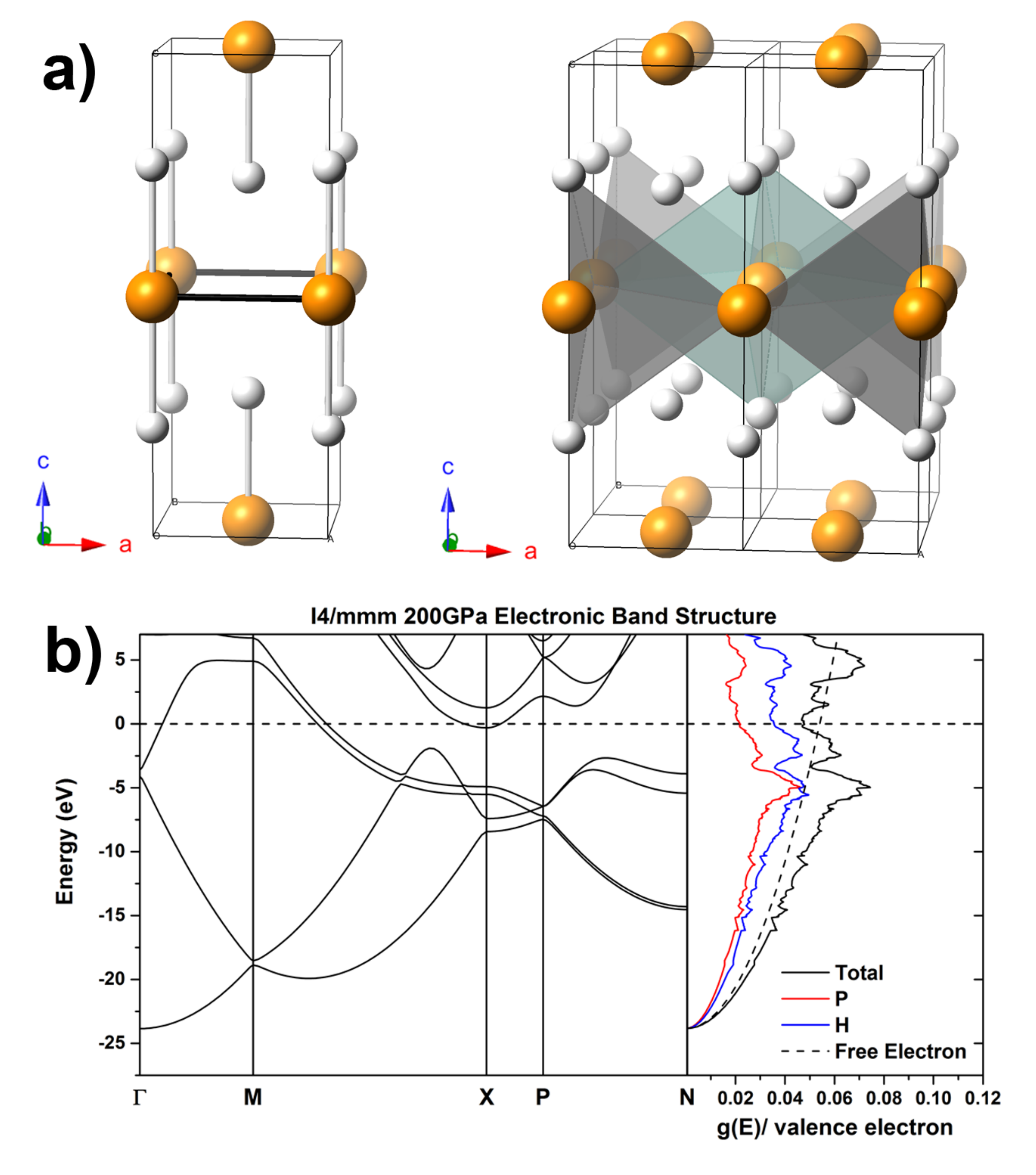}
\end{center}
\caption{Same as Fig.\ \ref{fig:Figure3} except for $I4/mmm$ PH$_2$ at 200~GPa. $\measuredangle$PPP=90$^{\circ}$, $\measuredangle$HPP=90$^{\circ}$ and $\measuredangle$HPH=180$^{\circ}$. In the figure on the top right a 2$\times$2$\times$1 supercell is shown.
\label{fig:Figure4}}
\end{figure}

Above $\sim$160~GPa two structurally related PH$_2$ phases, 2FU-${C2/m}$ (Fig.\ \ref{fig:Figure3}) and $I4/mmm$ (Fig.\ \ref{fig:Figure4}), become more stable than 5FU-${C2/m}$. The differences in the enthalpies of these two compounds at 150 and 200~GPa are less than 1~meV/atom, but in between these pressures the non-ZPE corrected enthalpy of the $I4/mmm$ phase is slightly lower than that of 2FU-${C2/m}$. The existence of nearly isoenthalpic hydrogen-rich phases under pressure in the DFT calculations is not unprecedented. We have previously found nearly isoenthalpic but distinct structures in our theoretical studies of hydrides containing an alkali metal or alkaline earth metal under pressure. This includes two Li$_5$H phases that were computed as being more stable than Li and LiH above 50~GPa \cite{Zurek:2012a}, five unique CsH$_3$ structures that were preferred over CsH and H$_2$ between 30-150~GPa \cite{Zurek:2012g} and three BaH$_6$ phases that had nearly the same enthalpies around 100~GPa \cite{Zurek:2012n}.

In both the ${C2/m}$ and $I4/mmm$ phases of PH$_2$ each phosphorous is octahedrally coordinated by four phosphorous atoms in the equatorial positions, and two hydrogen atoms in the axial positions. The difference between the two structures is that whereas in $I4/mmm$ all of the atoms assume the ideal octahedral angles, in 2FU-${C2/m}$ the hydrogens are canted with respect to a line that lies normal to the phosphorous square net. At 100~GPa the canting angle is negligible ($<$0.1$^{\circ}$), and the P-P and P-H bond lengths in the two structures are nearly identical (Table\ \ref{tab:Distances}). At 160 and 200~GPa, however, the canting angle increases to 9.8 and 10.1$^{\circ}$, respectively. 
Even though the nearest neighbor P-P/P-H distances in $I4/mmm$ are slightly smaller than in 2FU-${C2/m}$ at 160 and 200~GPa, the volume of the former is somewhat larger than that of the latter due to the tilting of the octahedra. The more efficient packing of the octahedra between the layers, which is a consequence of the canting, can most easily be seen by comparing the P-P distances between the phosphorous atoms comprising different layers. At 100~GPa the separation is identical, 2.983~\AA{}, resulting in comparable volumes. By 160~GPa, however, this distance becomes 2.757 and 2.695~\AA{} for $I4/mmm$ and 2FU-$C2/m$, respectively, giving rise to the sudden difference in volume of 1.2\% between the two phases. The $PV$ contribution to the enthalpy is smaller in the more compact 2FU-${C2/m}$ phase, but the electronic contribution more strongly favors the more symmetric $I4/mmm$ structure (see the SI), such that $I4/mmm$ is preferred by a few meV/atom for $P\sim$155-195~GPa. By 200~GPa, however, the two phases become isoenthalpic. 

Similar to 5FU-${C2/m}$, regions of high ELF in 2FU-$C2/m$ and $I4/mmm$ are found only along the P-P and P-H bonds and the Bader charges in Table \ref{tab:Distances} are indicative of charge transfer from phosphorous to hydrogen. Both of these phases are good metals with $g(E_F)$= 0.045-0.048~eV$^{-1}$/valence electron at 200~GPa, which is close to the value computed for a free electron gas with the same bandwidth, 0.054~eV$^{-1}$/valence electron. Their DOS plots display many features in common with the DOS calculated for 5FU-${C2/m}$: they are indicative of H/P hybridization and the character at the Fermi level is primarily hydrogen $s$-like, with substantial contributions from phosphorous $p$-states. Phonon calculations revealed that these two structures are dynamically stable at 100, 150, and 200~GPa.

Because of the structural similarity between the 2FU-$C2/m$ and $I4/mmm$ phases, we wondered if they may be connected via a low energy pathway. The NEB method was employed to find the barrier and transition state between the two structures. The computed barrier was $\sim$270~meV/atom at 200~GPa, implying that these phases likely lie within two distinct wells on the potential energy surface. The transition state (whose coordinates are given in the SI) can best be described as the average of the two PH$_2$ phases. Within the NEB method, the positions of the individual atoms within the starting and ending structures must match as closely as possible to avoid large geometric changes that may result in unrealistic barriers. In order to do this the $I4/mmm$ structure has to be reduced down to its primitive triclinic cell and then a 2 FU supercell in the crystallographic $b$-direction  direction must be constructed. In the transition state structure the perfect 90$^{\circ}$ square phosphorus nets described above have been distorted to diamonds with internal angles of 67 and 113$^{\circ}$ resulting in nearest and second nearest neighbor P-P distances of 1.885 and 2.581
\AA{} (c.f.\ Table 1), and the distance separating the P-P layers is 2.941~\AA{}, which can be compared to 3.017 and 2.963~\AA{} for the $I4/mmm$ and $C2/m$ phases, respectively. Each phosphorus is [2+4] coordinate to hydrogen. Similar to the two end point structures, the shortest two P-H distances fall between 1.45 and 1.50~\AA{}, which is slightly longer than the distance of a P-H bond in phosphine at ambient conditions, 1.42~\AA{} \cite{CRC:1966}. The remaining four P-H distances fall within 1.7-1.8~\AA{}, and are different for each phosphorous atom in the P1 symmetry 2FU transition state. 

We also wondered if tilting the octahedra to different angles would yield other structures not found in our evolutionary runs whose enthalpies were similar to $I4/mmm$ and $C2/m$. In order to test this hypothesis we relaxed structures constructed to have canting angles of 45$^{\circ}$, as well as geometries where hydrogen atoms were placed at random positions within the phosphorous lattice at 200~GPa. No unique low enthalpy structures were found. In addition, a closer inspection of the evolutionary runs showed that structures similar to those we constructed by hand had been generated and optimized during the EA searches, but their enthalpies were significantly higher than 5FU-$C2/m$, 2FU-$C2/m$ and $I4/mmm$ PH$_{2}$. 

Because the ZPE may be large for extended systems containing light elements, we computed the enthalpic differences between these three PH$_2$ phases including the ZPE corrections at pressures where they were dynamically stable. The open symbols in Fig.\ \ref{fig:Figure1}(c) show that the inclusion of the ZPE makes the $I4/mmm$ and 2FU-$C2/m$ structures significantly more stable than 5FU-$C2/m$ at 150~GPa, with the $C2/m$ being preferred by 6~meV/atom. Because the 2FU-$C2/m$ and $I4/mmm$ phases are structurally similar, so is their phonon DOS, and their ZPE at 150~GPa is nearly identical. The phonon DOS of the 5FU-$C2/m$ phase, on the other hand, shows that the frequencies associated with the hydrogen vibrations above 2000~cm$^{-1}$ are larger in comparison to these two phases, resulting in a significantly larger ZPE at 150~GPa, which leads to the destabilization of this phase when compared to the other two structures. At 200~GPa the ZPE corrections favor the $I4/mmm$ phase such that its zero-point corrected enthalpy is 25~meV/atom lower than that of 2FU-$C2/m$. The reason for this is that the $C2/m$ structure has more high frequency modes above 1500~cm$^{-1}$ (see Fig.\ \ref{fig:H5Ifigure2}). Thus, our results present another example of how the ZPE of phases with light elements can affect their relative enthalpies. It is beyond the scope of this work to consider anharmonic effects to the (free) energies (and $T_c$), but it may be that these also influence the order of stability of the phases studied herein. To determine if these three PH$_2$ structures could potentially contribute to the superconductivity observed by Drozdov and co-workers in their compression of phosphine up to 207~GPa, their superconducting properties were investigated in further detail, as described in the following section.



\subsection{Superconducting Phases of PH$_2$}

Above 5~GPa superconductivity has been observed in black phosphorous, the most stable allotrope at ambient conditions. The experimentally measured $T_c$ depends on the path taken in the pressure/temperature phase diagram. Recent experiments, and DFT calculations using the  Allen-Dynes modified McMillan equation, both showed that for the simple cubic phase of phosphorous $T_c$ decreases with increasing pressure \cite{Karuzawa:2002a,Cohen:2013a}. Computations revealed that the decrease in $T_c$ above 30~GPa could be explained by the increase in the phonon frequencies, and a $\mu^*$ of 0.18 yielded a $T_c$ of 5.8~K at 70~GPa \cite{Cohen:2013a}, which agrees well with the experimentally measured $T_c$ of 4.3~K at 100~GPa \cite{Karuzawa:2002a}. At higher pressures, up to 160~GPa, experiments suggested that $T_c$ decreased below 4~K \cite{Karuzawa:2002a}. In comparison, the computational methodology used herein coupled with a $\mu^*$ of 0.18 resulted in a $T_c$ of 6.4~K for simple cubic phosphorous at 100~GPa, and 
slightly higher temperatures for smaller values of the Coulomb pseudopotential, see Table \ref{tab:Tc}. We have also computed the parameters entering the Allen-Dynes modified McMillan equation for simple hexagonal phosphorous at 150 and 200~GPa, so that they may be compared with those calculated for the aforementioned PH$_2$ phases. As Table \ref{tab:Tc} shows, the average logarithmic frequency, $\omega_\text{log}$, of phosphorous increases with pressure and the electron-phonon coupling, $\lambda$, decreases quenching the superconductivity within the simple hexagonal phase.

\begin{figure}\begin{center}\includegraphics[width=0.8\columnwidth]{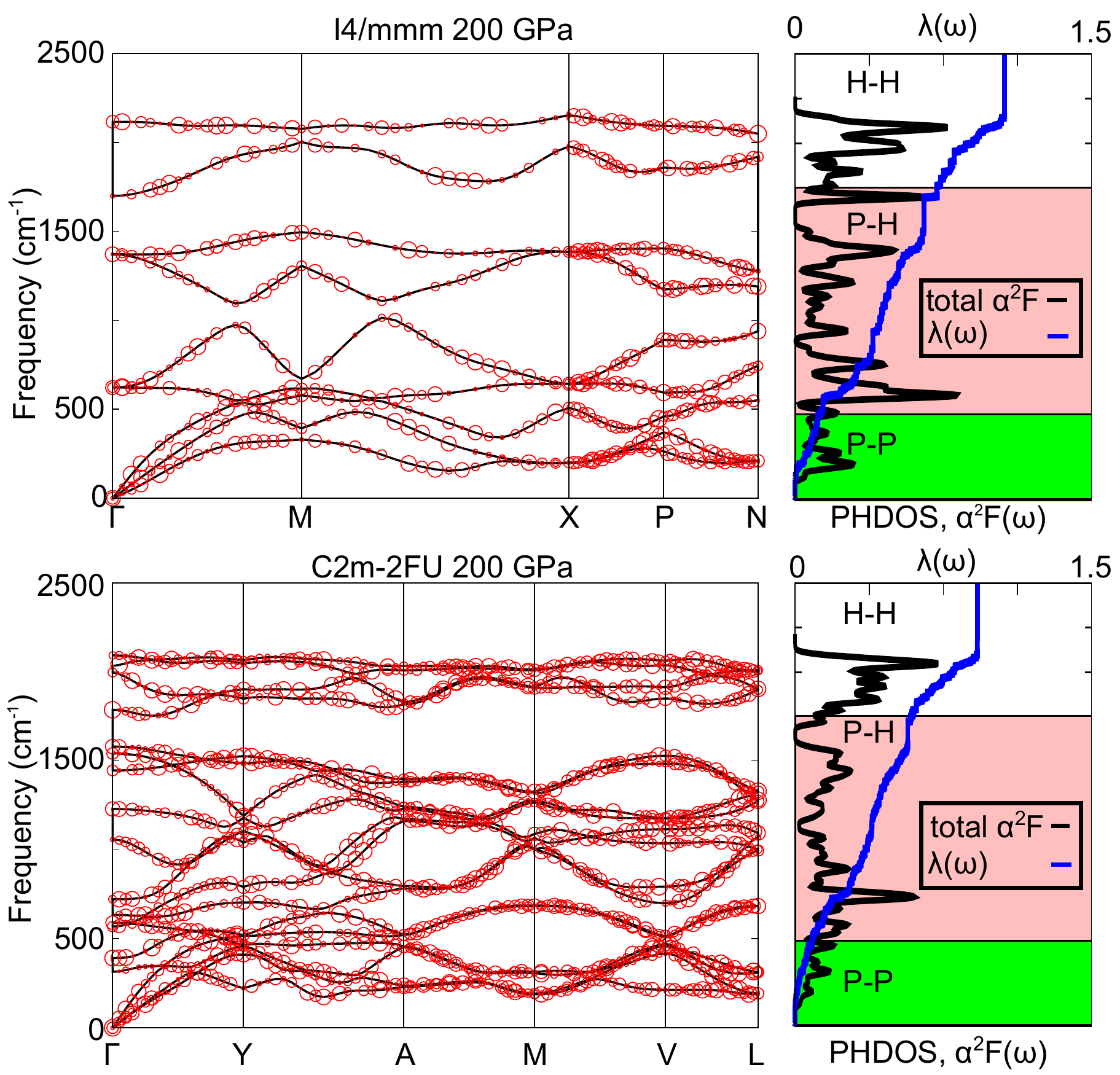}
\end{center}\caption{Phonon band structure, Eliashberg spectral function, $\alpha^2F(\omega)$, and the electron-phonon integral, $\lambda(\omega)$, for 2FU-$C2/m$ and $I4/mmm$ PH$_2$ at 200~GPa. Circles indicate the phonon linewidth with a radius proportional to the strength. The highlighted sections of the $\alpha^2F(\omega)$ plots show the division into vibrational modes that are comprised of phosphorus vibrations (P-P, green), motions of hydrogen and phosphorus atoms (P-H, pink), and mainly hydrogen vibrations (H-H, white). The contribution towards $\lambda$ for these divisions are 0.09 (9.3$\%$), 0.50 (54.6$\%$), and 0.33 (36.1$\%$) for 2FU-$C2/m$ and 0.13 (12.8$\%$), 0.58 (54.9$\%$), and 0.34 (32.3$\%$) for $I4/mmm$, for the green, pink, and white divisions, respectively.
\label{fig:H5Ifigure2}}\end{figure}

How would the incorporation of H atoms within the phosphorus lattices, in the most stable PH$_2$ geometries, affect the $\omega_\text{log}$, $\lambda$ and $T_c$? To answer this question we computed the superconducting properties of the previously discussed PH$_2$ phases, at pressures where they were found to be dynamically stable, and the results are given in Table \ref{tab:Tc}. Fig.\ \ref{fig:H5Ifigure2} provides the phonon band structure and linewidths, Eliashberg spectral function, and plots the dependence of $\lambda$ on the frequencies for the 2FU-$C2/m$ and $I4/mmm$ phases at 200~GPa (analogous plots can be found in the SI for all of the other $T_c$ values computed herein). 
In pure compressed hydrogen it has been proposed that $\mu^*$  can range from 0.08 to 0.089\cite{Ashcroft:1997,Ceperley:2011}, whereas values of 0.18\cite{Cohen:2013a} have been employed for simple cubic phosphorus at 100 GPa. We therefore list critical temperatures computed for $\mu^*$ of 0.1 and 0.18 in Table \ref{tab:Tc}. However, because the generally accepted value for the Coulomb pseudopotential is within 0.1 to 0.13, the text quotes $T_c$ values computed for a $\mu^*$ of 0.1.
\begin{table*}
\centering
\caption{Electron-phonon coupling parameter ($\lambda$), logarithmic average of phonon frequencies ($\omega_{\text{log}}$) and estimated superconducting critical temperature ($T_c$) for values of the Coulomb pseudopotential ($\mu^*$) of 0.1 and 0.18 for simple cubic (S.C.) phosphorous, simple hexagonal (S.H.) phosphorous, and the 5FU-$C2/m$, 2FU-$C2/m$, and $I4/mmm$ PH$_2$ phases at various pressures.}

\begin{tabular}{c c c c c c c c}
\hline
System & Pressure (GPa) & $\lambda$ & $\omega_{\text{log}}$ (K)  & $T_c^{\mu^*=~0.1}$ (K) & $T_c^{\mu^*=~0.18}$ (K) \\
\hline
S.C. P & 100 & 0.66 & 521.7 & 15.9 & 6.4 \\
S.H. P & 150 & 0.21 & 575.6 & 0.00 & 0.00 \\
S.H. P & 200 & 0.13 & 671.8 & 0.00 & 0.00 \\
5FU-$C2/m$ & 100 & 1.05 & 655.1 & 49.0 & 32.1 \\
5FU-$C2/m$ & 150 & 1.00 & 798.1 & 55.5 & 35.2 \\
2FU-$C2/m$ & 200 & 1.04 & 1026.5 & 75.6 & 49.2 \\
$I4/mmm$ & 150 & 0.86 & 946.2 & 50.6 & 28.3 \\
$I4/mmm$ & 200 & 1.13 & 851.6 & 70.4 & 48.0 \\\hline
\end{tabular}\label{tab:Tc}
\end{table*}

Because of the structural similarities between $C2/m$ and $I4/mmm$ at 150 GPa, their critical temperatures should be nearly identical, and $T_c$ was calculated as being 51 K for the $I4/mmm$ structure. The aforementioned canting of the octahedra results in geometric changes between these two structures at higher pressures, which could potentially affect the superconducting properties. Indeed, at 200~GPa the electron phonon coupling of $I4/mmm$ was calculated as being somewhat larger than for $C2/m$,  but $\omega_{\text{log}}$ was nearly 175~K larger for $C2/m$ because the vibrational modes arising primarily from hydrogen atoms are, on average, found at higher frequencies in this phase. The canting results in a larger number of hydrogen atoms that are found within a distance of 2~\AA{} of a given hydrogen in $C2/m$, thereby increasing the frequencies associated with their vibrations. For example, whereas $I4/mmm$ has four H-H distances at 1.486~\AA{}, in $C2/m$ the H-H measures are 1$\times$1.497~\AA{}, 1$\times$1.507~\AA{}, 2$\times$1.585~\AA{}, and 2$\times$1.775~\AA{}. Because of the opposing effects of $\omega_{\text{log}}$ and $\lambda$ on the critical temperature, the $T_c$ values computed for the $C2/m$ (76~K) and $I4/mmm$ (70~K) phases at 200~K are very similar despite their structural differences. 

At 200~GPa the most significant contribution to $\lambda$ in the 2FU-$C2/m$ and $I4/mmm$ PH$_2$ systems stems from vibrational modes between 500-1750~cm$^{-1}$ that involve hydrogen and phosphorus motions. The electron phonon coupling arising from modes above 1750~cm$^{-1}$, which consist primarily of the motion of hydrogen atoms with a small share from the phosphorous atoms, are also substantial components of the total $\lambda$. Vibrations below 500~cm$^{-1}$, which are the result of phosphorus motions, play a small, but non-negligible role in the electron-phonon coupling. The computed $T_c$ of the PH$_2$ phases are significantly larger than for monoatomic phosphorus because hydrogen increases both $\omega_\text{log}$ and $\lambda$. The electron phonon coupling is in-line with the values typically computed for compressed hydrogen-rich phases, 0.5-1.6\cite{Livas:2012a,Gao:2008a,Tanaka:2007a}, but smaller than the values of 2.19 at 200~GPa and 2.69 at 150~GPa computed for $Im\overline{3}m$ H$_3$S \cite{Cui:2014a} and CaH$_6$ \cite{Ma:2012f}, respectively. The phonon linewidths reveal that unlike CaH$_6$ \cite{Ma:2012f}, where the electron phonon coupling was derived primarily from a single vibrational mode, a plethora of modes contribute towards $\lambda$ in PH$_2$. 

The superconducting critical temperatures we compute for the $I4/mmm$ and 2FU-$C2/m$ phases of PH$_2$ are somewhat lower than the experimentally measured value of 103~K at 207~GPa \cite{Troyan:arxiv}. However, it may be that the  $T_c$ measured by Drozdov and co-workers results from a mixture of phases. For example, recent experimental work has suggested that the decomposition of LiH under pressure in a diamond anvil cell may lead to the formation of layers with different LiH$_n$ stoichiometries\cite{pepin}. The experiments were inspired by theoretical predictions\cite{Zurek:2009}, and observables computed via DFT calculations were employed to aid the interpretation of the experimental results. Pepin and co-workers suggested that under pressure Li diffuses into the diamond anvil cell forming a LiH$_6$ layer at the diamond/sample interface, and an LiH$_2$ layer at the LiH$_6$/LiH interface. Such mechanisms may also be important for the pressure induced decomposition of phosphine, and it is only via comparison of the computed experimental observables for specific phases with the results obtained experimentally that one can uncover which phases are formed under pressure, and the mechanisms underlying their formation. A feedback loop between experiment and theory is integral to advance our understanding of high pressure phenomena.

The decrease in $T_c$ with pressure of all of the PH$_2$ phases considered herein correlates with the markedly lower critical temperature, 30~K, measured by Drozdov and co-workers at 83~GPa. In future work we will focus on predicting the structures of PH$_n$ phases at pressures where vdW interactions may be important, and interrogate their bonding and superconducting properties.


\section{Conclusions}

Recent experiments revealed that when PH$_3$ is compressed to 207~GPa, it becomes superconducting below 103~K  \cite{Troyan:arxiv}. Our density functional theory (DFT) calculations have shown that at pressures of 100, 150 and 200~GPa PH$_3$ is thermodynamically unstable with respect to decomposition into the elemental phases, as well as PH$_2$ and H$_2$. Based upon the computed enthalpies other reactions that may occur under pressure are: PH$_3+$H$_2\rightarrow$PH$_5$ (100 and 150~GPa) and PH$_2\rightarrow\frac{1}{2}$H$_2$+PH (100~GPa).  \emph{A priori} crystal structure prediction has been used to identify three PH$_2$ phases that are dynamically stable in the pressure range of 100-200~GPa. All of these are calculated to be superconducting via the Allen-Dynes modified McMillan equation, suggesting that like H$_3$S \cite{hns}, PH$_2$ may be another hydrogen-rich Bardeen-Cooper-Schrieffer (BCS)-type superconductor. 

Two PH$_2$ structures with $C2/m$ and $I4/mmm$ symmetry were computed to have a superconducting critical temperature, $T_c$, of 76~K and 70~K, respectively, at 200~GPa.  Between 150-200~GPa the non-ZPE (zero point energy) corrected enthalpies of these phases differed by only a few meV/atom, and their structures both consisted of square nets of phosphorous atoms. In the $I4/mmm$ structure the phosphorous atoms were octahedrally coordinated, with two hydrogen atoms in the axial positions. In the $C2/m$ phase the hydrogens are slightly canted from the ideal octahedral angles allowing the distance between the planes of phosphorous atoms to decrease, and reducing the volume of the structure. A five formula unit phase with $C2/m$ symmetry (5FU-$C2/m$), which consists of 1-D PH$_{3}$-PH-PH$_{2}$-PH-PH$_{3}$ oligomers, was also identified in our evolutionary searches, and it has the lowest non-ZPE corrected enthalpy at 100~GPa. At this pressure it's $T_c$ is computed to be 49~K. Upon decreasing the pressure, the $T_c$ of all of the PH$_2$ phases decreased. 

Our results provide another example of how pressure can lead to the formation of compounds with stoichiometries and properties that would not be predicted based upon our experience at 1~atm. Comparison of the observables computed for these PH$_2$ phases with results obtained experimentally will unveil which phase, or mixture of phases, give rise to the superconducting properties observed by Drozdov and co-workers \cite{Troyan:arxiv}.

\paragraph{Supplementary Information Available:} Details of the structure searches, band structures, structural parameters, DOS plots, phonon DOS plots, absolute energies in Hartree, phonon convergence plots, and Nudged Elastic Band results. This material is available free of charge via the Internet at http://pubs.acs.org.

\section{Acknowledgements}

We acknowledge the NSF (DMR-1505817) for financial, and the Center for Computational Research (CCR) at SUNY Buffalo for computational support. A.S.\ acknowledges financial support from the Department of Energy National Nuclear Security Administration under Award Number DE-NA0002006, and  E.Z.\ thanks the Alfred P. Sloan Foundation for a research fellowship (2013-2015). A.S. also acknowledges Daniel Miller for his help with using the Nudge Elastic Band Method.


\providecommand*{\mcitethebibliography}{\thebibliography}
\csname @ifundefined\endcsname{endmcitethebibliography}
{\let\endmcitethebibliography\endthebibliography}{}

\end{document}